\documentclass{WileyMSP-template}

\usepackage{hyperref}

\begin{document}

\pagestyle{fancy}

\title{Interfacial Tension Hysteresis of Eutectic Gallium-Indium}
\maketitle

\author{Keith D. Hillaire$^{1,*}$}  
\author{Praneshnandan Nithyanandam$^{2,*}$}
\author{Minyung Song$^2$} 
\author{Sahar Rashid Nadimi$^{2,3}$} 
\author{Abolfazl Kiani$^{2,4}$}  
\author{Michael D. Dickey$^2$}  
\author{Karen E. Daniels$^{1}$}

\begin{affiliations}
$^1$Dept. of Physics, NC State University, Raleigh, NC

$^2$Dept. of Chemical and Biomolecular Engineering, NC State University, Raleigh, NC

$^3$Physical Science Department, Bakersfield College, Bakersfield, CA 

$^4$Department of Chemistry and Biochemistry, California State University, Bakersfield, CA

$^*$Joint first authors

\end{affiliations}


\date{\today}

\begin{abstract}
When in a pristine state, gallium and its alloys have the largest interfacial tensions of any liquid at room temperature.
Nonetheless, applying as little as 0.8 V of electric potential across eutectic gallium indium (EGaIn) placed within aqueous NaOH (or other electrolyte) solution will cause the metal to behave as if its interfacial tension is near zero.
The mechanism behind this phenomenon has remained poorly understood because NaOH dissolves the oxide species, making it difficult to directly measure the concentration, thickness, or chemical composition of the film that forms at the interface. In addition, the oxide layers formed are atomically-thin.
Here, we present a suite of techniques which allow us to simultaneously measure both electrical and interfacial properties as a function of applied electric potential, allowing for new insights into the mechanisms which cause the dramatic decrease in interfacial tension.  A key discovery from this work is that the interfacial tension displays hysteresis while lowering the applied potential.  
We combine these observations with electrochemical impedance spectroscopy 
to evaluate how these changes in interfacial tension arise from chemical, electrical, and mechanical changes on the interface, and close with ideas for how to build a free energy model to predict these changes from first principles.
\end{abstract}

\section{Introduction}

\begin{figure}
    \centering
    \includegraphics[width=0.6\linewidth]{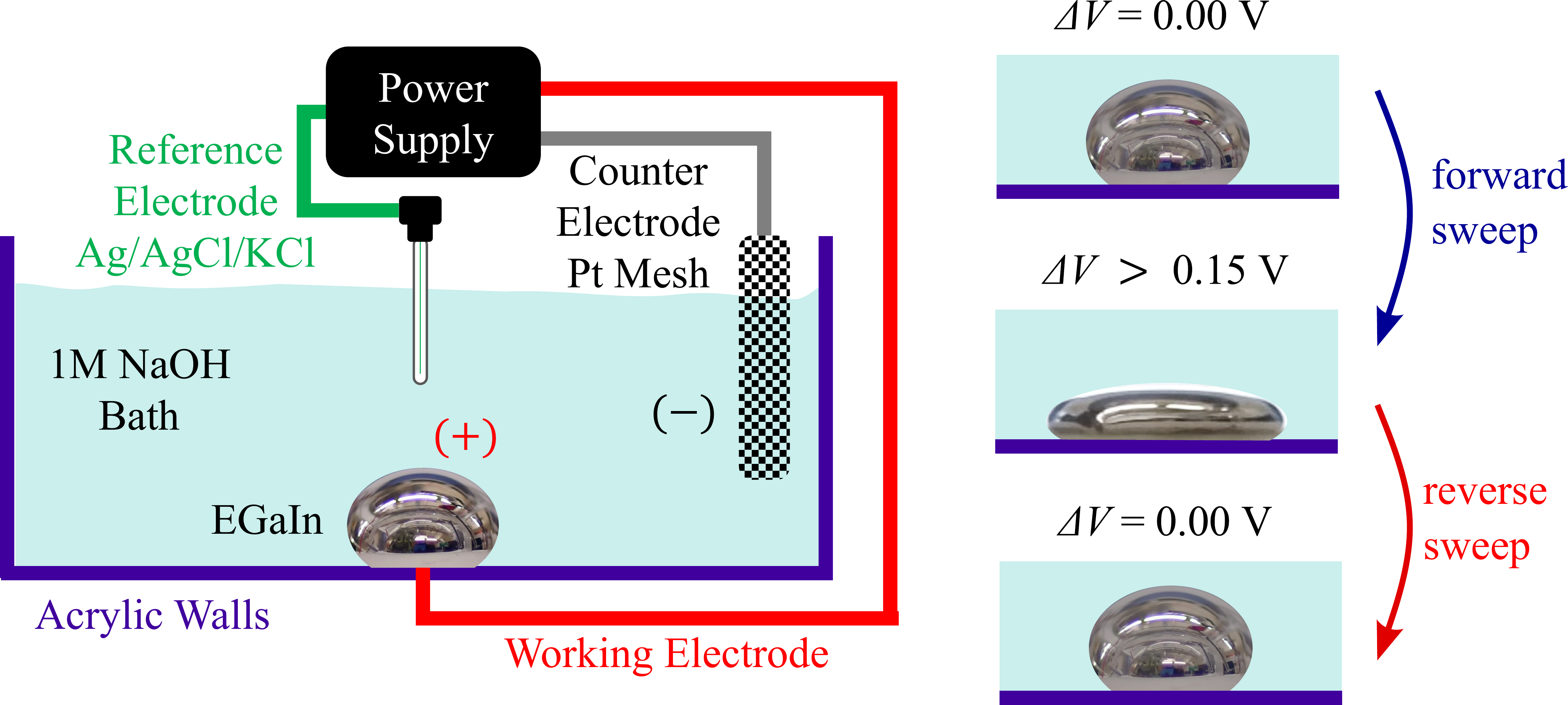}
    \caption{(a)A 100 $\mu$L sessile droplet of EGaIn is in placed in a 1M NaOH solution (light blue background), with the three-electrode configuration shown schematically.  At open circuit potential ($\Delta V = 0$~V), the liquid metal has  high interfacial tension, but flattens above $\Delta V \approx 0.15$~V due to a dramatic drop in interfacial tension. When the applied voltage is returned to $\Delta V = 0$~V, the higher interfacial tension returns and the droplet retracts to its original shape. }
    \label{f:SessileSetup}
\end{figure}

A suite of desirable properties --- electrochemically tunable interfacial tension down to negligible values, high thermal and electrical conductivity, low viscosity, non-toxicity --- make gallium-based liquid metals particularly promising for use in soft and reconfigurable devices \cite{dey_microfluidically_2016,eaker_electrowetting_2017,khan_giant_2014,joshipura_stretchable_2017,chen_liquid_2018,vallem_energy_2021,wang_electrochemically_2021}, including pumps \cite{tang_liquid_2014}, actuators, \cite{yu_discovery_2018}, valves \cite{berre_electrocapillary_2005}, and RF devices \cite{gough_continuous_2014} by injecting the liquid metals into microfluidic channels \cite{khoshmanesh_liquid_2017,kim_multiaxial_2008,dickey_eutectic_2008,khan_recapillarity_2015,jung_highly_2015}. Of particular interest is the room-temperature liquid eutectic gallium indium (EGaIn).
The reversible and highly-effective electrochemical tunability of its interfacial tension is remarkably different from conventional, surfactant-based approaches which only modestly decrease the interfacial tension (by $\approx 20-50$~mN/m) \cite{de_aguiar_interfacial_2010} and are typically difficult to remove \cite{liu_switchable_2006}.
The electrochemical methods further surpass the effectiveness of electrocapillarity, \cite{eaker_electrowetting_2017,tang_liquid_2014,berre_electrocapillary_2005,gough_rapid_2015,t_mannetje_stickslip_2013,cha_thermal_2016}, whereby an applied electric potential modifies the electric double layer on the surface of the metal in an electrolyte and thereby lowers the interfacial tension without changing the chemistry \cite{vallem_energy_2021}. A schematic of the electrochemical method is shown in Figure~\ref{f:SessileSetup}. When a voltage is applied to a droplet and then removed, the droplet flattens and then returns to its original shape.
 
Utilizing electrochemical oxidation of the surface of EGaIn within an electrolyte provides rapid  and reversible control of interfacial tension over an enormous range ($\approx$ 500 mN/m at 0 V to nearly 0 mN/m at $\approx 0.8$~V) \cite{khan_giant_2014,eaker_oxidation-mediated_2017}, well beyond what is achievable by conventional methods for controlling surface effects.
The dramatic effects of this tunability generate surprising fluid behaviors through exploiting classic fluid instabilities under applied electric potentials: destabilizing droplets into fractal spreading patterns \cite{chen_liquid_2018,eaker_oxidation-mediated_2017}, or stabilizing thin liquid streams \cite{song_overcoming_2020,he_noncontact_2022}. 

While Figure~\ref{f:SessileSetup} illustrates the interfacial properties reversing after a complete voltage cycle, there is hysteresis present during intermediate steps. By examining the detailed dynamics under forward and reverse voltage sweeps, we can use these intermediate steps to test hypotheses about the specific electrochemical processes which underlie the observations. 
The observed decrease in interfacial tension has been speculated to arise from surfactant-like oxide(s) forming by electrochemical reactions, with the observed drop in interfacial tension coinciding with the onset of oxide formation \cite{khan_giant_2014, eaker_oxidation-mediated_2017}, as has also been observed that the interfacial tension of liquid aluminum decreases with the oxygen content of its environment \cite{gheribi_temperature_2019}.
Further studies suggested the oxide species may additionally create compressive forces at larger applied voltages \cite{eaker_oxidation-mediated_2017}, but the precise mechanisms have remained poorly understood.
In this paper, we aim to critically evaluate the possible chemical, electrical, and interfacial mechanisms that lead to the dependence of the interfacial tension on applied electric potential by comparing impedance spectroscopy and interfacial tension measurements, and highlight future avenues for study.

\section{Results and Discussion}

\subsection{Cyclic voltammetry on sessile droplets}

\begin{table}[b]
\centering
\begin{tabular}{|c | l|}
\hline
{\bf Symbol} & {\bf Definition} \\ \hline \hline
$\Delta V$ & difference from open circuit voltage \\ \hline
$V_\mathrm{OC}$ & open circuit voltage ($\Delta V = 0$) \\ \hline
$V_\mathrm{P}$	& onset voltage of passivation \\ \hline
$V_\mathrm{SQ}$	& onset voltage of quadratic decrease in tension \\ \hline
$V_\mathrm{L}$	& onset voltage of linear decrease in tension \\ \hline
	\end{tabular}
\caption{Table of voltage abbreviations}
\label{t:voltages}
\end{table}

Using a sessile (stationary) droplet of 100 $\mu$L of EGaIn (Ga 75.5\%,  In 24.5\%) in a NaOH solution, as shown in Figure~\ref{f:SessileSetup}, we simultaneously measure (1) the interfacial tension $\gamma$ from image analysis of the droplet shape during anodic oxidation and (2) the electrical current $I$ using a potentiostat with a Ag/ AgCl/ KCl reference electrode to maintain an applied electric potential $\Delta V$. Each experimental run consists of a cycling $\Delta V$ between the open circuit potential $V_\mathrm{OC}$, a maximum applied potential $V_\mathrm{max}$, and back to  $V_\mathrm{OC}$ at a quasistatic rate of 10 mV/s. 
The potentiostat records the  current $I$ passing between the working electrode and  the counter electrode; to determine the current density $J = I/a$ considered the area $a$ of the droplet's upper, exposed surface. The surface area was measured from the detected edge of the droplet, integrated under the assumption of a radially symmetric droplet. Details of these methods are given in  \S\ref{s:methods}.

We image the droplet from the side; by synchronizing this video with step controls on the DC power supply, we measure $\gamma(\Delta V)$. As shown in Figure~\ref{f:ForwardRegimes}, we observe four clear transitions to which we assign names. 
The first occurs at the  sharp rise in current density $J$, which we associate to the open circuit potential $V_\mathrm{OC}$, located at  $-1.53$~V measured with respect to the Ag/AgCl/KCl electrode. We assign this value to be $\Delta V = 0$, and all voltages in this paper are measured relative to it. The second transition occurs at the passivating voltage $V_\mathrm{P} \approx (0.10 \pm 0.01) \, \mathrm{V}$, at which the current $J$ sharply drops to a plateau two orders of magnitude lower. The third transition, at $V_\mathrm{SQ} \approx (0.15  \pm 0.01) \, \mathrm{V}$, is associated with the start of an approximately quadratic shape with $\Delta\gamma(\Delta V) \propto \Delta V^2$, followed by the fourth transition to an approximately linear trend above $V_\mathrm{L} \approx (0.27  \pm 0.01) \, \mathrm{V}$.
Within this forward sweep, we observe that the interfacial tension  is approximately constant ($\gamma_0 \approx (500  \pm 15) \, \mathrm{mN/m}$) over the range $ V_\mathrm{OC} < \Delta V < V_\mathrm{SQ}$, after which a two-stage decrease occurs. The transition at $V_\mathrm{P}$ is further demarcated by the observation that for experiments in which the same NaOH solution was reused multiple times, an additional dip in $\gamma(\Delta V)$ was observed at this point.

\begin{samepage}
\begin{figure}
  \centering
  \includegraphics[width=0.45\columnwidth]{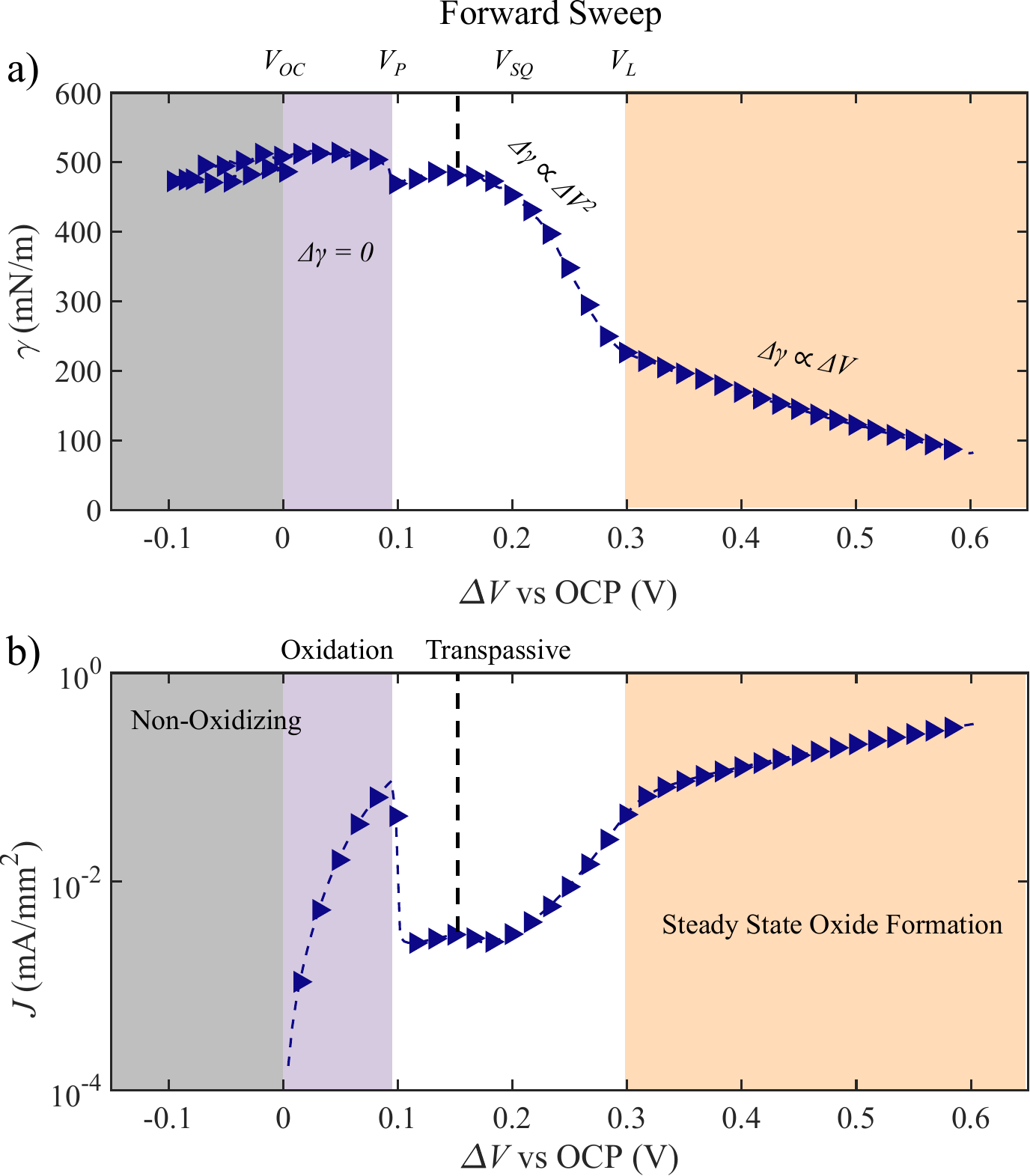}
  \caption{Typical forward voltage sweep experiment (10 mV/s sweep rate), with voltage shown relative to the open circuit potential $V_\mathrm{OC}$ and $\Delta V_\mathrm{max} = 0.6$~V. Voltage-dependence of the (a) interfacial tension  $\gamma$ and (b) current density $J$ are shown in parallel, with the four  electrochemical regimes described in the text marked as colored bands:  \textit{non-oxidizing}, \textit{oxidation}, \textit{transpassive}, and \textit{steady state oxide formation}. The dashed line corresponds to the transition between  diffusive and inductive transpassive behavior. A summary of the voltage variables and their subscripts appears in Table~\ref{t:voltages}.}
  \label{f:ForwardRegimes}
\end{figure}

\begin{figure}
  \centering
  \includegraphics[width=0.45\columnwidth]{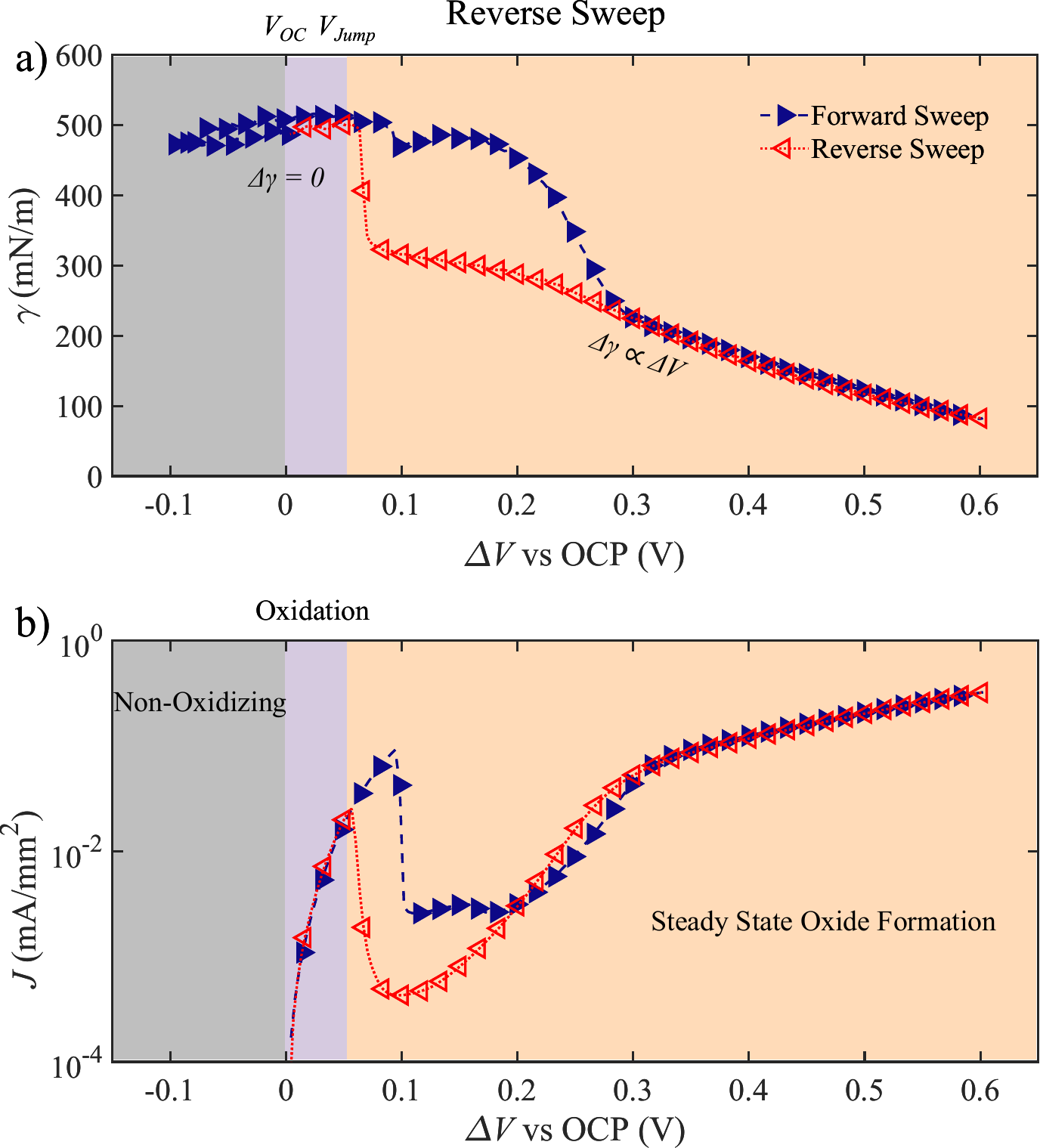}
  \caption{Typical hysteresis experiment (10 mV/s sweep rate), with voltage shown relative to the open circuit potential $V_{OC}$  and $\Delta V_\mathrm{max} = 0.6$~V. Forward (voltage increasing) sweep is plotted as  dark blue right-facing triangles, and the reverse (voltage decreasing) sweep as  red left-facing triangles.
  Voltage-dependence of the (a) interfacial tension  $\gamma$ and (b) current density $J$ are shown in parallel,  with only three of the electrochemical regimes observed --- \textit{non-oxidizing}, \textit{oxidation}, and \textit{steady state oxide formation} --- as defined by the behavior of $\gamma$  and $J$ in the reverse sweep. The transpassive regime is not observed on the reverse sweep. Note that the purple curves are repeated from Figure~\ref{f:ForwardRegimes}.}
  \label{f:ReverseRegimes}
\end{figure}

\end{samepage}

For the reverse voltage sweep (from $\Delta V_\mathrm{max}$ back to zero), these transitions are modified. As shown in Figure~\ref{f:ReverseRegimes}, we observe hysteresis at intermediate voltages. 
We identify a voltage $V_\mathrm{jump} \approx (0.06  \pm 0.01) \, \mathrm{V}$ above which the interfacial tension $\gamma$ and current density $J$ are history-dependent. This voltage is below the passivating transition at $V_\mathrm{P}$ observed during the forward sweep, and is characterized by sharp increases in both $\gamma$ and $J$ as the voltage is lowered. Once the system returns below $V_\mathrm{jump}$, the interfacial tension returns to $\gamma_0 \approx (500  \pm 15) \, \mathrm{mN/m}$.

Within the hysteretic regime, $\gamma(\Delta V)$ no longer follows the quadratic shape observed during the forward sweep, and instead follows a linear rise that remains on the same the linear trend seen for $\Delta V > V_\mathrm{L}$. This suggests that the quadratic drop in $\gamma$ seen in the forward sweep does not arise from purely electrocapillarity effects ($\Delta\gamma_\mathrm{EC} = -\frac{C}{2} \Delta V^2$) because those would not display any hysteresis.

\subsection{Electrochemical impedance spectroscopy on sessile droplets}

\begin{figure}
  \centering
\includegraphics[width=0.6\columnwidth]{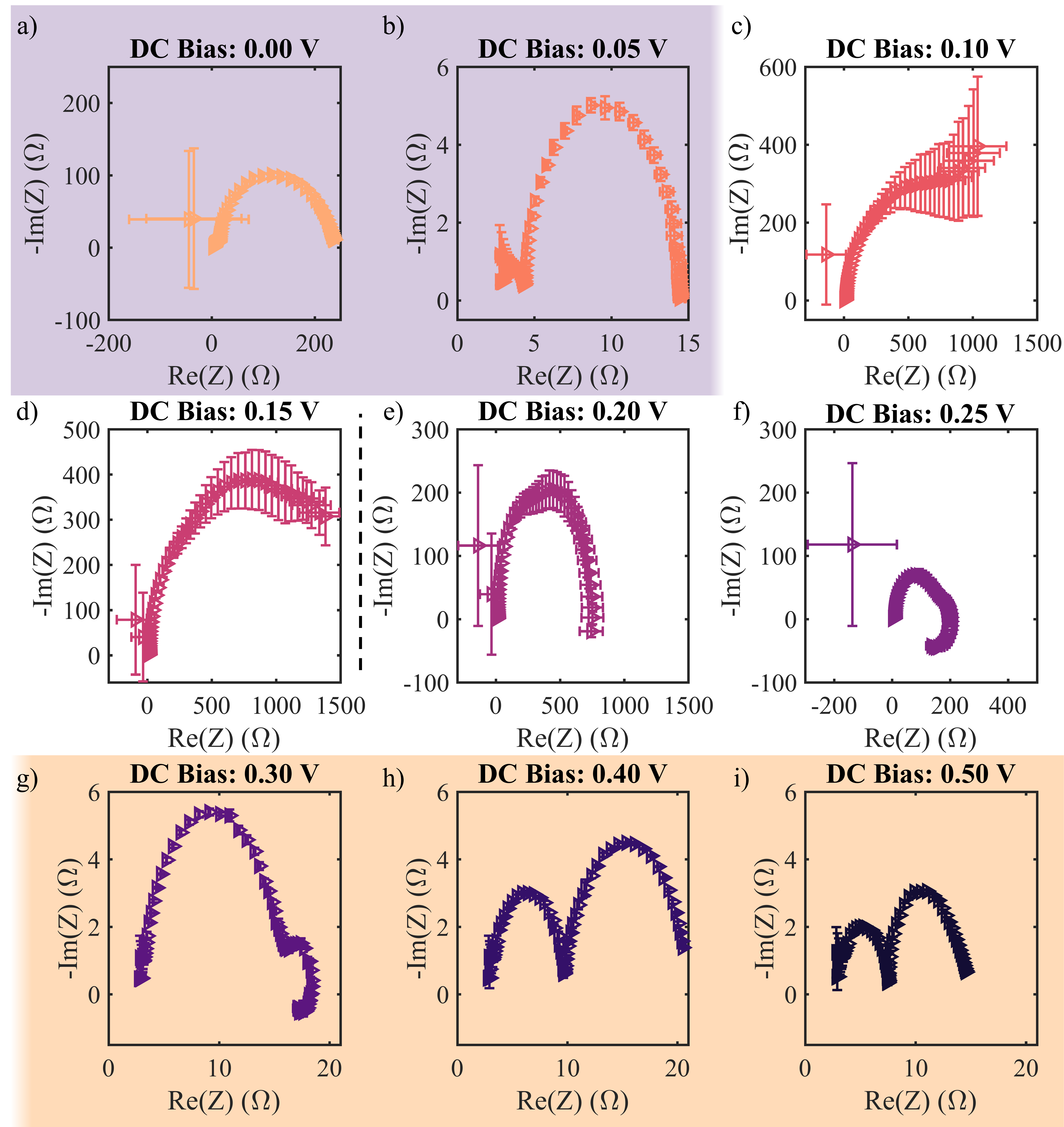}
  \caption{Nyquist diagrams of the electrochemical impedance spectroscopy at various DC biases during a forward potential sweep; each data point at a specific DC bias is the average of all six runs (both H2L and L2H) and the error bars are the standard deviation among them. (a,b) Low voltage \textit{oxidation} regime, with purple background. (c-f) Medium voltage  \textit{transpassive} regime, with white background (g-i) High voltage \textit{steady state oxide formation} regime, with orange background. Background colors correspond to those used in Figure~\ref{f:ForwardRegimes}, and the dashed line between (d) and (e) corresponds to the same line drawn in Figure~\ref{f:ReverseRegimes}, the transition between  diffusive and inductive transpassive behavior.}
  \label{f:EIS05avg}
\end{figure}

Using the same sessile droplet apparatus, again with 100 $\mu$L EGaIn droplets in a 1M NaOH solution, we additionally performed electrochemical impedance spectroscopy (EIS) measurements \cite{vivier_impedance_2022} and compared these results to the $\gamma(\Delta V)$ and $J(\Delta V)$ measurements presented above.  All EIS experiments were performed with a constant DC bias ($\Delta V$)  superimposed with an AC voltage of amplitude 5 mV at frequencies ranging from 100 mHz to 100 MHz; the DC bias was varied from  0 V to 0.5 V, corresponding to the range of the forward/backward cyclic voltammetry sweeps.
Each DC bias was run six times, three times from high to low frequency (H2L), and three times from low to high frequency (L2H); the average of all six runs is plotted in Figure~\ref{f:EIS05avg}, and the error bars are the standard deviation among the runs. We further note that the data for the two frequency sweep directions match each other within the level of fluctuations seen within individual sweeps, validating these measurements.
The EIS data was verified using the Kramer-Kronig test, as shown in the Supplemental Material, and details of the EIS methods are given in  \S\ref{s:methods}.

\subsection{Interpretation of regimes}

Using the EIS data of the droplet at each $\Delta V$, combined with observed transitions in $\gamma(\Delta V)$ and $J(\Delta V)$, we construct physical interpretations of the regimes marked by colored bands in Figures~\ref{f:ForwardRegimes} and \ref{f:ReverseRegimes}.  At low current and voltage, the system is \textit{non-oxidizing}. This regime is not the focus of our paper, but appears in our plots wherever $J < 0 \,\mathrm{mA/mm}^2$. Here, the surface of the EGaIn is being reduced rather than oxidized, since $J \leq 0$ when $\Delta V \leq V_\mathrm{OC}$. This observation is what allows us to reference all of our data to $V_\mathrm{OC}$, and define $V_\mathrm{OC} \equiv 0 \, \mathrm{V}$. 

During the lowest voltages of our forward sweeps ($ 0 \, \mathrm{V} < \Delta V <  V_\mathrm{P} \approx 0.1 \, \mathrm{V}$), $J$ is positive and increases monotonically with $\Delta V$ (see Figure~\ref{f:ForwardRegimes}b). 
As we increase the applied potential above the open circuit potential, the surface of the droplet becomes \textit{oxidizing} and we give this name to the regime shown in purple. In the Nyquist plot shown in Figure~\ref{f:EIS05avg}ab, we observe that $\mathrm{Re}(Z)$ decreases compared to the values measured at $\Delta V = 0$. This  indicates a decrease in charge transfer resistance, and causes the electrical current $J$ to  increase as shown in Figure~\ref{f:ForwardRegimes}b. In addition, this is a charged interface, and exhibits a capacitance (or psuedocapacitance). Within this regime, $\gamma$ is approximately constant (see the initial flat region in Figure~\ref{f:ForwardRegimes}a):  if the surface contains any ions (as suggested by the impedance data), they are not surface-active.  In addition, the dissolution rate of the gallium oxide species is at least as large as the oxidation rate.

This regime ends for $V > V_\mathrm{P}$, at which point  $J$ drops by two orders of magnitude (see Figure~\ref{f:ForwardRegimes}b). We interpret this as the minimum voltage at which the oxidized gallium species first coats the entire surface of the droplet, creating a passivating layer that strongly suppresses additional electrochemical reactions. We call this regime above $V_\mathrm{P}$ the \textit{transpassive} regime, as it occurs after the breakdown of passivation. 

Within this regime (between $V_\mathrm{P}$ and $V_\mathrm{SQ}$), we observed small changes in $J$ depending on the extent to which the NaOH solution was reused, but there were no changes in $\gamma$ observed when using fresh 1M NaOH.
The reuse of solution presumably creates conditions in which  the concentration of dissolved  gallium becomes large enough to cause a dip in $\gamma$ around $V_\mathrm{P}$. Fortuitously, these experiments allow us to conclude that the oxidation species must not be changing the chemical potential of the Ga--NaOH$_\mathrm{aq}$ interface enough to observe the change until the concentration of Ga dissolved in NaOH$_\mathrm{aq}$ becomes significant. Therefore, we believe the oxidation species is an adsorbed gallium hydroxide such that the total number of gallium atoms and hydroxide ions at the interface does not change above and below $V_\mathrm{P}$. Gallium oxide dissolves into Ga(OH)$^{-}_\mathrm{4}$, which is stable in its dissolved form in NaOH \cite{diakonov_gallium_1997}, and therefore the concentration of Ga(OH)$^{-}_\mathrm{4}$ in the NaOH solution increases with each run of the experiment. Once the concentration of Ga(OH)$^{-}_\mathrm{4}$ is high enough, the dip in $\gamma$ at $\Delta V \approx V_\mathrm{P}$ appears in our data. This hypothesis is supported by the Nyquist plots: the impedance data decays into a linear tail, which implies the interface is diffusive \cite{pletcher_instrumental_2011,orazem_electrochemical_2008}. Future direct measurements of ion concentration (\textit{e.g.} Ga$^{3+}$) would allow for testing the validity of these interpretations.

Interestingly, we observe a small local maximum in both $\gamma$ and $J$ near the middle (0.15 V, dashed vertical line) of the \textit{transpassive} regime, which indicates that the oxide species on the surface of the metal changes to a different species in the vicinity of this potential \cite{bard_electrochemical_2001}. This is a much smaller effect than the initial oxidation peak in $J$ that delineates the \textit{oxidation} regime. The large oxidation peak likely corresponds to the adsorption of gallium hydroxides, while this secondary transpassive peak arises from the formation of gallium oxide, either Ga$_\mathrm{2}$O$_\mathrm{3}$ or GaOOH according to the literature \cite{chung_electrochemistry_2013}.
As the system moves into this \textit{transpassive} regime, the composition of the oxide film starts to change; this can be identified by the Nyquist plots changing from diffusive (Figure~\ref{f:EIS05avg}c,d) to inductive (panels e,f).
This also coincides with changes in the interfacial tension (see Figure~\ref{f:ForwardRegimes}). Such a change in oxide 
has been observed in anodic dissolution of mild steel, suggesting presence of an adsorbed layer of FeOH associated with transpassivation
\cite{li_impedance_1996}. 

From the secondary transpassive peak onward to higher $\Delta V$, marked by a dashed black line in the Figure~\ref{f:ForwardRegimes}, we observed that  $\gamma$ decreases quadratically (proportional to $\Delta V^2)$. If this oxide layer has approximately constant capacitance in this subregime (as will be measured in \S\ref{s:model}), then  the energy density of the surface layer (J/m$^2$, equivalent to interfacial tension N/m) due to the resulting charge separation would be proportional to $\Delta V^2$, as is observed here.  Note, however, that this $\gamma$-dependence only appears during the forward sweep, and not the reverse sweep. This suggests that any capacitive  charge separation is not reversible, and is perhaps only metastable. As will be discussed in \S\ref{s:model}, this explanation is preferred over electrocapillary explanations due to the anomalously large values of the capacitance observed.

Care is required when interpreting  the  Nyquist plots within the  quadratic subregime ($V_\mathrm{SQ} < \Delta V < V_\mathrm{L}$) due to a few caveats. Within our data, we observe an increase in both the real and imaginary components of the impedance for sequential runs, a phenomenon known as drift. This arises from changes to the surface of the working electrode that are irreversible within the timescale between runs of the experiment \cite{ziino_investigation_2020}.
Since we believe the interface transitions from gallium hydroxide to gallium oxide within this regime, such drift  is to be expected. We also observe an induction loop that crosses the $\mathrm{Im}(Z) = 0$ line, which is likely due to the droplet physically oscillating, generating new surface area during the frequency sweep. 

Finally, above $V_\mathrm{L}$, the system switches multiple behaviors all at the same voltage: $J$ becomes an exponential function of $\Delta V$, $\gamma$ decreases linearly with $\Delta V$, and the Nyquist plots show a transition from having a single semicircle with an induction loop that crosses $\mathrm{Im}(Z) = 0 \, \Omega$ to having two semicircles with $\mathrm{Im}(Z) < 0$ for all frequencies measured. We interpret the $J$-dependence as arising from new reactions at the  metal-oxide interface, in which ions are driven through the surface layer \cite{macdonald_evaluation_1988}; we call this the \textit{steady state oxide formation} regime. In this interpretation, at  $\Delta V = V_\mathrm{L}$, the surface concentration $\Gamma_\mathrm{ox}$ has reached its maximum concentration for a single layer, and any new oxide now must form multiple layers. Such behavior is consistent with what has been observed for  aluminum in its \textit{steady state oxide formation} regime: that the amount of oxide shell material is proportional to the applied anodization potential \cite{mibus_dielectric_2013}, and that the decrease in $\gamma$ is due to stress in the oxide layer induced by $\Delta V$ \cite{haiss_surface_2001}. 
The Nyquist plots in this regime (see Figure~\ref{f:EIS05avg}g-i) all show two semicircles, an observation consistent with the system having two timescales as is typically seen for coatings of metals. We further observe that experiments in this regime are  highly reproducible, as the oxide is likely too stiff to support the oscillations that interfered with measurements at lower voltages. 

As shown in Figure~\ref{f:ReverseRegimes}, only three of the four regimes are present during a reverse sweep due to hysteresis. There is no longer a quadratic subregime, and the two branches re-merge only for $V_\mathrm{jump} < V_\mathrm{SQ}$.
The hysteresis implies that the oxide layer remains thick enough for the mechanisms of the \textit{steady state oxide formation} regime to remain dominant, possibly because the gallium oxide shell is insoluble within the timescale of these experiments. This would mean the surface species do not revert during the reverse sweep, and instead maintain the behavior observed in the steady state oxide regime.

\begin{figure}
  \centering
  \includegraphics[width=0.7\columnwidth]{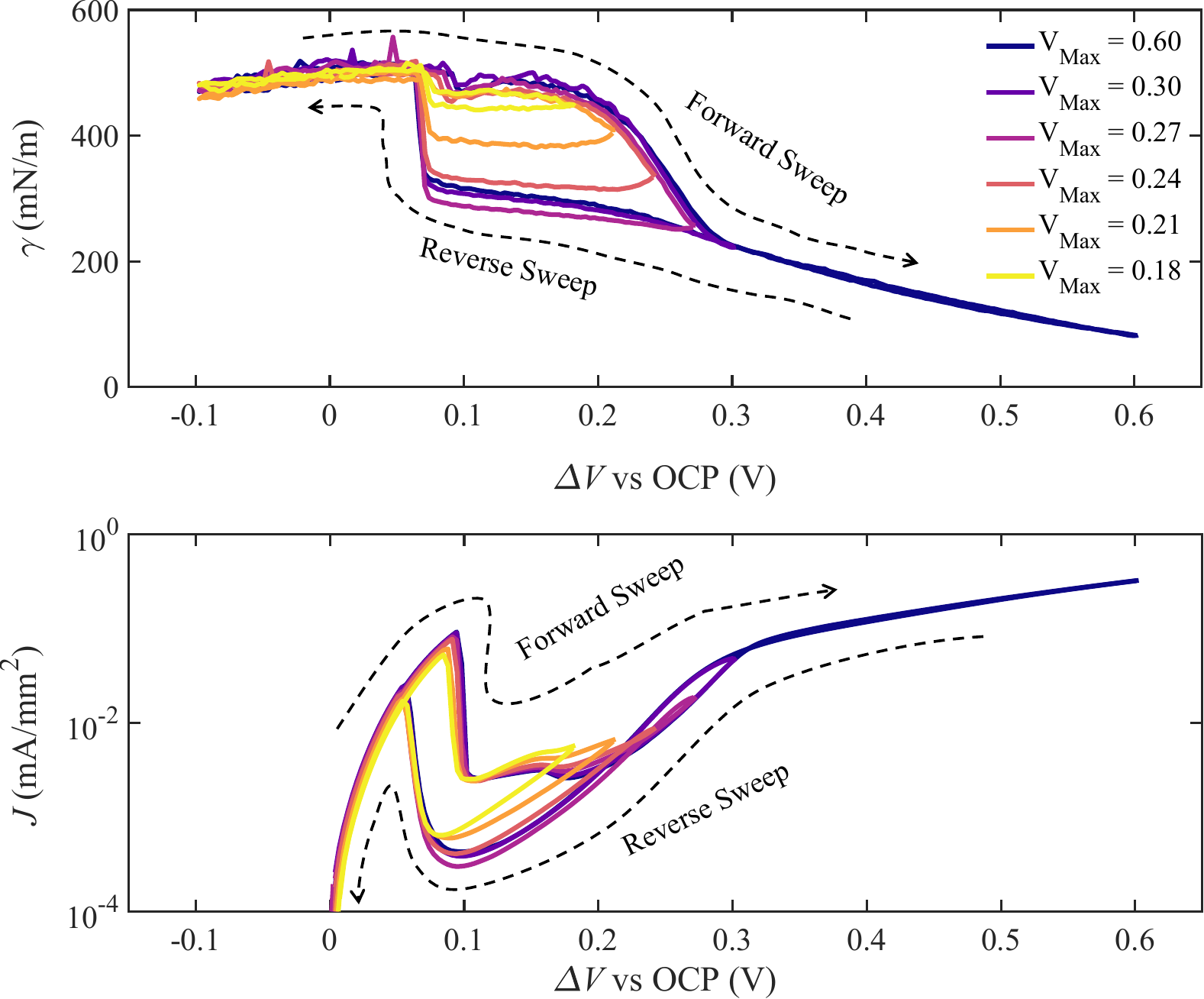}
  \caption{The observed hysteresis at a 10 mV/s sweep rate with various maximum voltages. Dark purple curves are at higher $V_\mathrm{max}$; light yellow curves are at lower $V_\mathrm{max}$.  The hysteretic transition is again visible (as in Figure~\ref{f:ReverseRegimes}) at $\Delta V_\mathrm{jump} \approx 0.06$~V; the curve  for  $\Delta V_\mathrm{max} = 0.6$~V is the same dataset. }
  \label{f:HystVmax}
\end{figure}

To test this hypothesis, we ran additional hysteresis experiments with different values of  $V_\mathrm{max}$, spanning  both the \textit{transpassive} and \textit{steady state oxide formation} regimes. The results are shown in Figure~\ref{f:HystVmax}. We observed that if $V_\mathrm{max} < V_\mathrm{L}$, then the hysteresis curve remains as presented above, while for  $V_\mathrm{max} < V_\mathrm{L}$ the hysteretic effects are reduced. In these less-hysteretic experiments, $\gamma$ remained constant upon reversal, again jumping back to the original branch at  $\Delta V = V_\mathrm{jump}$. This confirms that gallium oxide is likely insoluble for $\Delta V > V_\mathrm{jump}$ within the timescale of these experiments. {Notably, when $\Delta V < V_\mathrm{jump}$ the droplet maintains $\gamma_0$, likely due to the oxidation rate of the EGaIn dropping below the dissolution rate of the gallium oxide, after which the oxide is quickly removed from the surface. This is visible in Figure~\ref{f:HystVmax}. Since gallium oxide reduces the amount of hydroxide (OH$^-$ or Ga(OH)$_\mathrm{4}^{-}$) on the interface of the metal, it reduces the chemical potential  between the bulk EGaIn and the metal-electrolyte interface. This creates a decrease in the surface energy density, which is proportional to the oxide concentration.

We summarize all of the above, schematically,  in Figure~\ref{f:Cartoon}. During the forward sweep (increasing voltage, panel a), the EGaIn is bare from $V_\mathrm{OC}$ to $V_\mathrm{P}$, at which point the surface begins to be covered with adsorbed gallium hydroxide. From $V_\mathrm{SQ}$ to $V_\mathrm{L}$, the surface coverage transitions to being gallium oxide, with $\Delta\gamma \propto \Delta V^2$ an approximately constant capacitance across an oxide layer of approximately constant thickness (a monomolecular layer). Beyond $V_\mathrm{L}$, this oxide shell may become thicker, but such measurements are beyond the scope of this paper. In any case, $\Delta\gamma$ is now mainly arising due to a voltage induced stress in the oxide layer. During the reverse sweep (decreasing voltage, panel b), the oxide layer is gradually dissolved as the dissolution rate increases relative to the oxidation rate. At timescales longer than these experiments, this transient state might be eliminated if the sweep were slow enough. At $V_\mathrm{jump}$ the dissolution rate exceeds the oxidation rate by a large enough amount that the system returns to its original, bare state.

\begin{figure}
  \centering
  \includegraphics[width=\columnwidth]{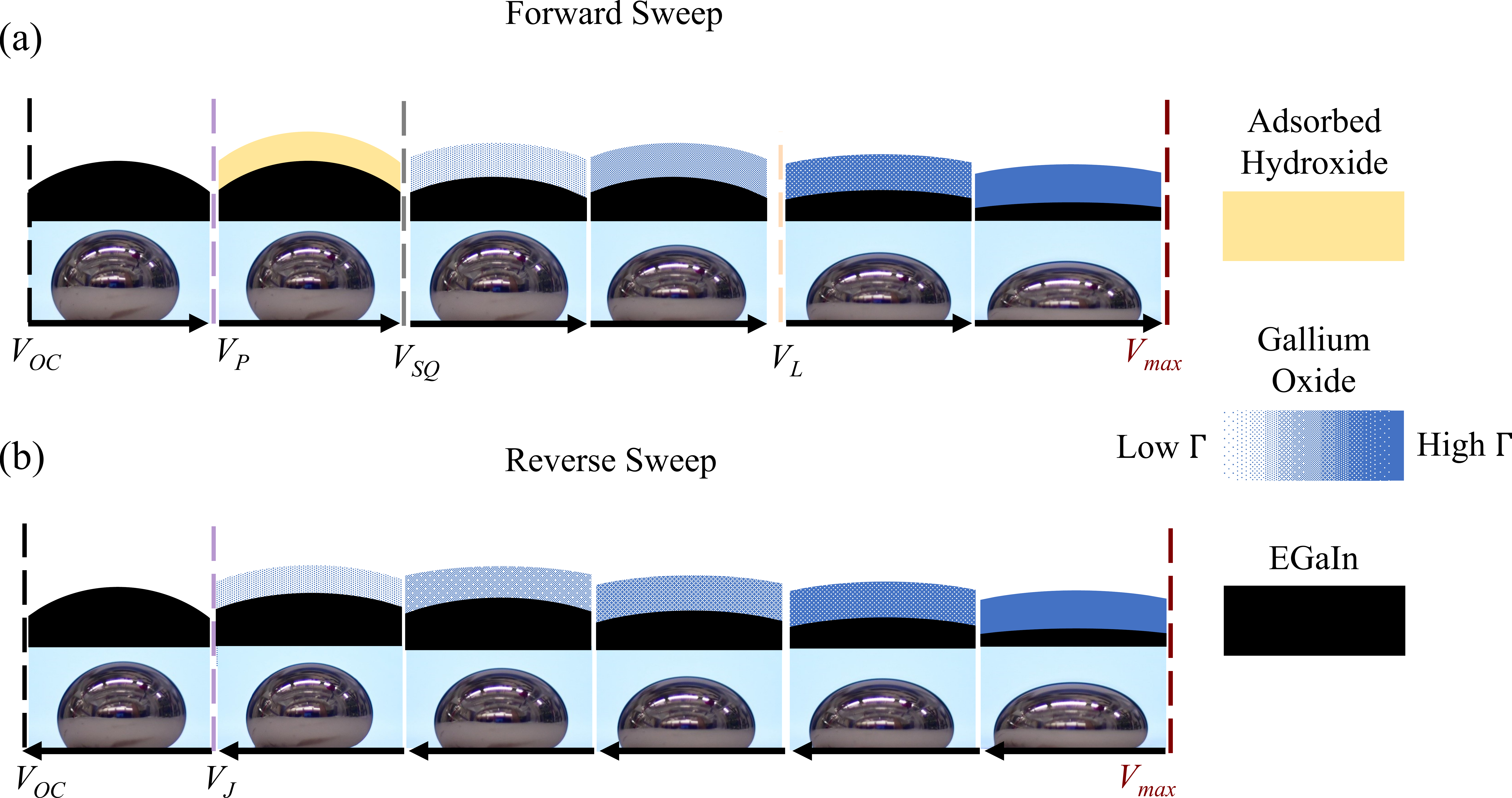}
  \caption{Cartoon of the proposed surface oxidation responsible for lowering the interfacial tension. Each EGaIn droplet is shown bare, with a light yellow layer representing adsorbed gallium hydroxide, or with a blue layer of changing $\Gamma$ (can either change due to surface density or oxide thickness).}
  \label{f:Cartoon}
\end{figure}

\begin{figure}
  \centering
  \includegraphics[width=0.7\columnwidth]{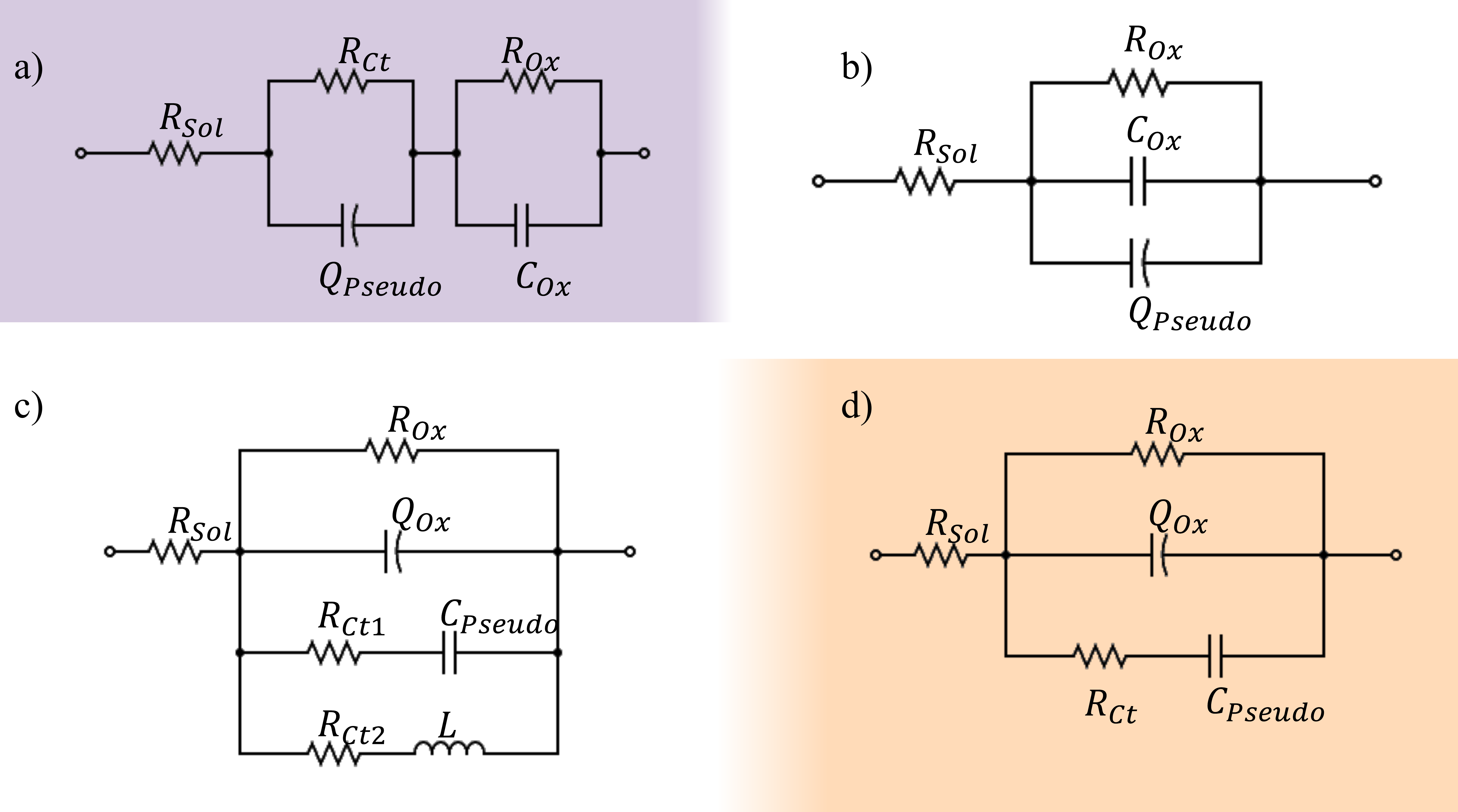}
  \caption{Equivalent circuits for the (a) oxidation, (b) early transpassive (diffusive), (c) late transpassive (inductive), and (d) steady state oxide formation regimes. Colors correspond to regimes marked in Figure~\ref{f:ForwardRegimes}.}
  \label{f:EISCircuits}
\end{figure}

\subsection{Modeling \label{s:model}}

Using our oxidation cartoon, we can create equivalent circuits (see Figure~\ref{f:EISCircuits}) for the different electrochemical regimes, and then use these to fit the Nyquist plots using the  capacitance and resistance values   as fit parameters. All circuits start with the resistance of the electrolyte  solution $R_\mathrm{sol}$, in series with additional effects associated to the particular regime. 

For the \textit{oxidation} regime (purple, see panel a),  we first include a circuit element to model the charge transfer resistance $R_\mathrm{ct}$, and a non-ideal pseudocapacitance $Q_\mathrm{pseudo}$; these describe the chemical reactions occurring at the interface, in parallel with each other. Second, there's an additional element with both the oxide resistance $R_\mathrm{ox}$ and capacitance, $C_\mathrm{ox}$, which describe the oxide itself.

For the \textit{transpassive} regime (white, see panels b and c), we include several elements in parallel: the resistance $R_\mathrm{ox}$ and capacitence $C_\mathrm{ox}$ of the  oxide layer, also in parallel with a pseudocapacitance $Q_\mathrm{pseudo}$ describing the reaction of a the species diffusing through and around the shell.
Once the voltage is above $V_\mathrm{SQ}$, it is necessary to adjust this circuit to account for the inductive behavior we see in the Nyquist plots. We have modeled these effects by making the oxide instead have a pseudocapacitance $Q_\mathrm{ox}$, and then adding  two parallel charge transfer paths, one describing the reaction with an ideal pseudocapacitance $C_\mathrm{pseudo}$ in series with a charge transfer resistance $R_{Ct1}$, and one describing the electrical induction $L$, caused by the changing area of the droplet in series with a charge transfer resistance $R_{Ct2}$.

For the  \textit{steady state oxide formation} regime (orange, see panel d), we remove the inductive behavior and place several elements in parallel: the resistance $R_\mathrm{ox}$ and pseudocapacitence $Q_\mathrm{ox}$ of the  oxide layer remain as before, but the additional charge transfer path is a charge transfer resistance $R_\mathrm{ct}$ in series with $C_\mathrm{pseudo}$, to model ions pushing through the oxide layer to reach the pure gallium interface.

\begin{figure}
  \centering
  \includegraphics[width=0.9\columnwidth]{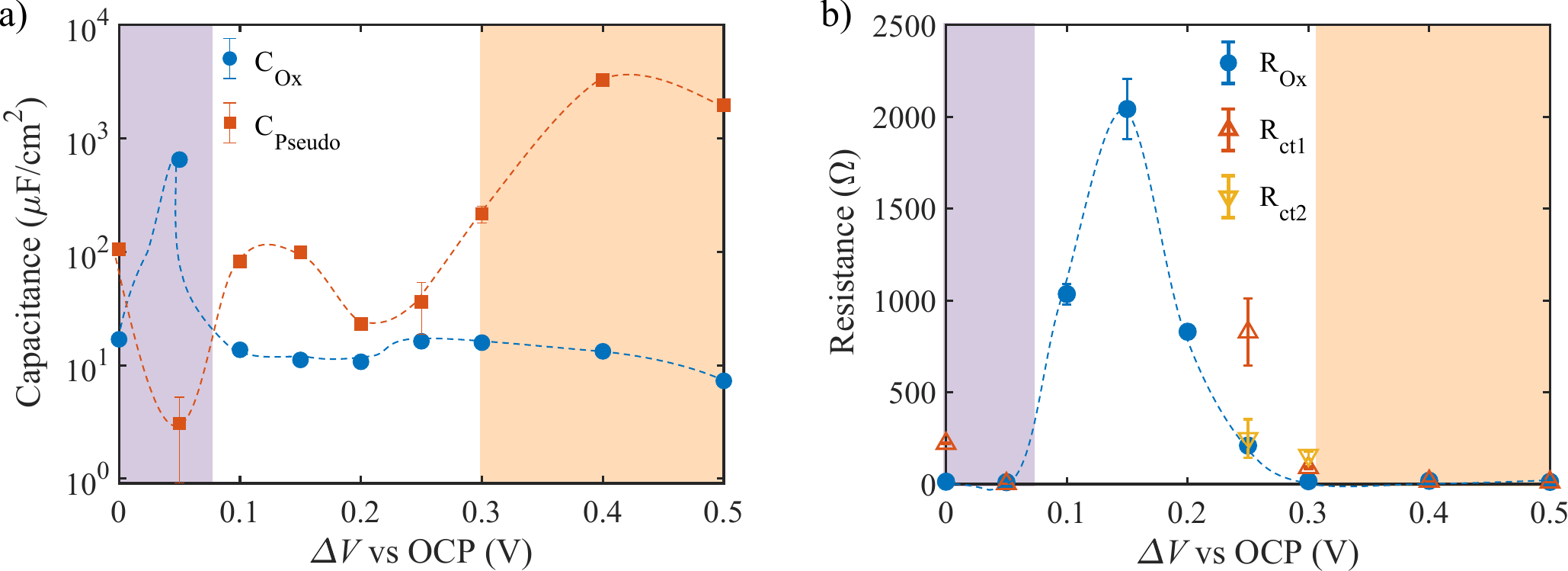}
  \caption{Parameters obtained by fitting the  L2H Nyquist plots to equivalent circuit models  (Figure~\ref{f:EISCircuits}) using a non-linear least square fit. Example fits are shown in the Supplemental Material. (a) Capacitance (blue cirlces) and pseudocapacitance (orange squares) and (b) oxide resistance (blue circles) and two charge transfer resistances (orange and yellow triangles). Dashed lines are to guide the eye and colored bands correspond to the regimes marked in Figure~\ref{f:ForwardRegimes}.}
  \label{f:EISFits}
\end{figure}

Figure~\ref{f:EISFits} evaluates the results of these simplified models when fitting data such as that shown in Figure~\ref{f:EIS05avg}. Panel a shows how $C_\mathrm{ox}$ and $C_\mathrm{pseudo}$ change with increasing applied voltage.  The oxide capacitance initially grows during the \textit{oxidation} regime, then drops and remains below $20 \, \mu \mathrm{F/cm}^{2}$ for higher voltages, while the pseudocapacitance remains around or below $100 \, \mu \mathrm{F/cm}^{2}$ until the \textit{steady state oxide formation} regime where it grows to over $1000 \, \mu \mathrm{F/cm}^{2}$. The resistance fits are shown in panel b: the oxide resistance grows until a maximum near $V_\mathrm{SQ}$, after which it drops again. 

These effective circuit parameters also permit a comparison to some idealized models for the interfacial tension, assuming that all capacitive elements store energy at the interface, and therefore correspond to surface energies \cite{hillaire_instabilities_2022}. 
In the quadratic subregime within the \textit{transpassive}, we observe $\Delta\gamma(\Delta V) \propto \Delta V^2$. 
If this were due solely to electrocapillarity effects, we would expect $\Delta\gamma(\Delta V) = C_\mathrm{EC}\Delta V^2$ \cite{grahame_electrical_1947}. For the observed changes in $\gamma$, this would require $C_\mathrm{EC} = (3200  \pm 50) \, \mu \mathrm{F/cm}^2$. As this value  is an order of magnitude larger than the values of $C_\mathrm{ox}$ we found by fitting the Nyquist plots in this regime (see Figure~\ref{f:EISFits}a), it is unlikely that the observed decrease in $\gamma$ can be accounted for solely by traditional electrocapillarity. Note that this value of $C_\mathrm{EC}$ is also an order of magnitude larger than $C_\mathrm{pseudo}$ measured in the same regime, but is on the same order as $C_\mathrm{pseudo}$ measured in the \textit{steady state oxide formation} regime.

We can understand the linear $\gamma$ regime (above $V_\mathrm{L}$ for the forward sweep, above $V_\mathrm{jump}$ for the backward sweep) in terms of the Gibbs adsorption isotherm equation. We consider $\Delta\gamma = -\sum{N_\mathrm{ox}/A}\Delta\mu_\mathrm{ox}$ \cite{butt_physics_2003}, where $N_\mathrm{ox}/A$ is the concentration of gallium oxide on the interface and $\Delta \mu_\mathrm{ox}$ is the difference in chemical potential between gallium oxide and pure gallium in 1M NaOH. While the values  for $N_\mathrm{ox}(\Delta V)$ and  $\Delta\mu_\mathrm{ox}$ are not measured here, we can assume $\Delta N_\mathrm{ox} \propto \Delta V$. Starting from an initially-bare metal interface, we integrate to find $N(\Delta V) = \frac{dN}{dV} \Delta V$, which predicts a functional form 
\begin{equation}
\Delta\gamma(\Delta V) = \frac{\Delta\mu_\mathrm{ox}}{A} \frac{dN}{dV} \Delta V
\end{equation}
for which we measure the prefactor to be
$\frac{\Delta\mu_\mathrm{ox}}{A} \frac{dN}{dV} = (59 \pm 4) \frac{\mathrm{nmol}}{\mathrm{V} \, \mathrm{cm}^2} \frac{\mathrm{kJ}}{\mathrm{mol}}$.
Unfortunately we do not have enough information to determine whether $N_\mathrm{ox}/A$ is changing by increasing the surface concentraion of the oxide, or by increasing the thickness of the oxide layer. If we did, we could use this value to estimate the relative permittivity of the oxide layer, which would help us determine the oxide species. Or if we knew the oxide species, we could determine if the oxide was growing in surface density or in thickness during this regime and we could compare this what we learned from the Nyquist plots. 

These models also allow us to understand some of the results presented in   Figure~\ref{f:EIS05avg}a-b, where the data are consistent with a decrease in charge transfer resistance $R_\mathrm{ct}$ in two voltage ranges: both from 0 and 0.0.5 V and also above 0.3 V. These are also the same two regimes for which holding the potential steady produces a steady state current that does not decay to zero. This implies that the oxide species that form on the surface due to electrochemical oxidation dissolve, and therefore do not accumulate on the surface beyond a steady state thickness. Such behavior is consistent with low charge transfer resistance. In the regime between 0.1 to 0.25 V, the current is very low due to passivation and thus, we do not report $R_\mathrm{ct}$ values in this range since resistance at the surface is dominated by the oxide ($R_\mathrm{Ox})$). Ongoing work in our laboratory suggests that the passivating species between $\sim 0.1$ to $0.3$~V may be GaOOH. In contrast, soluble species (gallate salts) form beyond the passivation regime (steady state oxide formation).

\section{Conclusion}

During this investigation into the dependence of the interfacial tension $\gamma$ on the applied electric potential $\Delta V$, we observed that $\gamma$ can be reproducibly tuned for $\Delta V > V_\mathrm{L}$, but for $V_\mathrm{P} < \Delta V < V_\mathrm{L}$ $\gamma$ depends on the history of the droplet, as shown in Figure~\ref{f:HystVmax}.
It is apparent that during the forward sweep (increasing $\Delta V$ from the open circuit potential to a more positive potential), there are three different behaviors of $\gamma$: $\gamma$ independent of $\Delta V$, $\gamma$ decreasing proportional to $\Delta V^2$, and $\gamma$ decreasing linearly with $\Delta V$. A small regime with a fourth behavior --- a small dip in $\gamma$ at $\Delta V \approx V_\mathrm{P}$ --- occurs only if the electrolyte solution is polluted with Ga(OH)$_\mathrm{4}^{-}$.

We have found that the regime where $\Delta\gamma \propto \Delta V^2$, the decrease in $\gamma$ is not likely due to electrocapillarity as (1) the capacitance required for this regime to be electrocapillarity is too large, and (2)  electrocapillarity should not show hysteresis. It is also  likely not due to a voltage-induced tangential stress in the oxide as changing the sweep direction below $V_\mathrm{L}$ leads to a $\Delta\gamma \approx 0$.
The $\Delta\gamma \propto \Delta V^2$ behavior occurs for values of $\Delta V$ beyond the second oxidation peak, and is therefore most likely due to the change in chemical potential of the interface due to the formation of an insoluble gallium oxide, either Ga$_\mathrm{2}$O$_\mathrm{3}$ or GaOOH. Future research could address making quantitative predictions for the prefactor, based on propoerties of the oxides present. 
The regime where $\gamma$ decreases linearly with $\Delta V$ can be attributed to a voltage-induced tangential stress in the gallium oxide layer.
In the reverse sweep, $\gamma$ increased linearly with decreasing $\Delta V$ for $\Delta V > V_\mathrm{jump}$, since the gallium oxide layers in insoluble at these potentials in the timescale of these experiments. Finally, 
$\gamma$ returns to the bare metal value for $\Delta V < V_\mathrm{jump}$, which appears to be due the oxidation rate of the EGaIn being much less than the dissolution rate of the oxide, which leads to the droplet reverting back to a bare metal at these electric potentials.

We have presented several studies which, when taken together, increase our understanding of possible mechanisms for the drop in interfacial tension with applied voltage. The hysteresis reported here has, to our knowledge, not previously been reported. The decrease in $\gamma$ for all oxidation potentials has  previously been assumed to be due to the concentration of gallium oxide on the surface of the metal. Once the exact molecular forms of the gallium hydroxides and gallium oxide are known, the surface concentrations can be determined as a function of $\Delta V$, it will be possible to build a free energy model of $\gamma$ as a function of $\Delta V$. However, directly observing the oxide at these electric potentials is difficult as all forms of gallium oxide are optically transparent, and so directly measuring the concentration of the oxide in this voltage range has proven a challenge. Measuring the rheology of anodizing EGaIn in 1M NaOH may help to better understand the mechanisms involved in the behavior of $\gamma(\Delta V)$, allowing the chemical and interfacial mechanisms to be distinguished. 

\section{Experimental Methods \label{s:methods}}

\subsection{Sessile Drop Technique}

The sessile drop technique is a technique in which a sessile (immobile) droplet is imaged from the side so that the shape of the droplet on a substrate can be directly measured. There are various ways the interfacial tension can be extracted from the image, from the one-component Zisman theory \cite{fowkes_contact_1964}, the two-component Owens/Wendt theory \cite{owens_estimation_1969}, to the three-component van Oss theory \cite{schrader_modern_1992}. For this work, we used calculations derived by Hutzler \textit{et al.} \cite{hutzler_simple_2018}: these are an  application of Morse-Witten theory  to sessile and pendant drops under a small force, originally derived to model the deformation of a droplet in an emulsion \cite{morse_droplet_1993}. It is valid for large values of $\gamma$, as we have here, and the  approximation is based on both the  maximum equatorial drop radius $L_x$, and the distance from the equator to the top of the droplet $L_z$.  As such, a key benefit is that it does not require measuring the contact angle between the EGaIn and the acrylic substrate; this is particularly important because droplets sometimes become pinned at their base. 

For a sessile droplet, $\gamma$ is calculated using the following equation:
\begin{equation}
    \gamma (L_z,L_x) \approx \Delta\rho g \frac{\ln{(2)}}{24} \frac{(L_z+L_x)^3}{|L_z-L_x|} (1 - 3c + 3c^2 - c^3)
\label{eq:hutzler}
\end{equation}
where the constant $c = \frac{1-\ln{(2)}}{\ln{(2)}} \frac{|L_z - L_x|}{L_z+L_x}$ is a correction due to the difference in the derived formula and the values  measured by Hutzler \textit{ et al.} \cite{hutzler_simple_2018}.
A needle with an outer diameter of 1.26 mm was placed in the photos, directly above the droplet, as a reference to determine the number of pixels per mm. We calculated the area $a$ of the droplet using edge detection. Assuming radial symmetry, $a$ is calculated by rotating the edges around the z-axis, $a = \sum_i \pi(\frac{|x_\mathrm{i+1}+x_\mathrm{i}|}{2}) \Delta s_i$, where $x_i$ and $y_i$ are the positions of the edge pixels, and $\Delta s_i^2 = \Delta x_i^2 + \Delta y_i^2$. A sample analysis is shown in the Supplementary Material (Figure~\ref{f:DropArea}).

\subsection{Cyclic Voltammetry}

To oxidize the liquid metal, we used a three electrode setup consisting of a working electrode, a counter electrode, and a reference electrode as shown in Figure~\ref{f:SessileSetup}. For the  majority of the experiments, we used a Biologic potentiostat; the EGaIn droplet acted as the working electrode.
A thin copper wire, which was connected to the power supply, was pushed through the bottom of an acrylic substrate, and a 100 $\mu$L of EGaIn was wet to the wire. The substrate was leveled so that the droplet would remain in place and in electrical contact with the wire during the course of the experiment.
A platinum mesh electrode was used for the counter electrode and was placed on the edge of the NaOH bath to minimize gradients in the electric field at the surface of the droplet. The Ag/AgCl/KCl reference electrode (2M KCl) was placed as close to the EGaIn as possible without touching the droplet at its widest spreading and being far enough away as not to interfere with the imaging of the droplet.

We tested several different sweep rates: 1 mV/s, 10 mV/s, 50 mV/s, 100 mV/s, and 300 mV/s, all from $-1.63$~V to $-0.93$~V, to determine the maximum sweep rate for which the droplet dynamics are significantly faster than the chemical reaction rates, so that the droplet remains near steady-state. The fastest of these sweep rates (300 mV/s) caused  the droplet shape to change so quickly that it was difficult to obtain an accurate measurement of $\gamma$. At 1 mV/s it was not possible to complete a full cycle during the Nikon D850's camera's available recording duration. Therefore, we selected 10 mV/s as the sweep rate, which is slow enough for the droplet dynamics to remain near steady-state. Sample data for a variety of sweep rates is shown in the Supplementary Material.

To compare the results of this experiment with other experiments of different shapes and sizes, $I$ was converted to current density $J$ by dividing out the exposed (upper) surface area $a$ of the droplet: $J = I/a$. At the base of the droplet, $J \approx 0 \, \mathrm{mA/mm}^2$ for the surface of the droplet in contact with the substrate, so it does not contribute to the current density. 

\subsection{Electrochemical Impedance Spectroscopy}

The EIS experiments used the same sessile drop design as the cyclic voltammetry experiments.
A copper wire was connected to a 100 $\mu$L EGaIn droplet, a platinum mesh was used for the counter electrode, and an Ag/AgCl/KCl electrode was used for the reference electrode.
The experiment was performed in a 1M NaOH bath using the Biologic potentiostat.
The EIS experiments were held at a constant DC bias with a 5 mV AC amplitude, applied from 100 mHz to 100 MHz with 12 points per decade. We observed that for a larger AC amplitude (10 mV), the results were less reproducible than at 5 mV. 

\section{Data availability}

Data is available for download at DataDryad: \url{https://doi.org/10.5061/dryad.2z34tmpsb}

\section{Acknowledgments}

The authors would like to thank Carmen Lee, Hangjie Jie,  William Llanos, Jeffrey Wong, Thomas Witelski, Mark Orazem, and Digby Macdonald for helpful discussions. We also like to thank the National Science Foundation for support under NSF DMR-1608097 (K.D.H. and K.E.D.) and CBET-1510772 (M.D.D.).

\bibliographystyle{MSP}
\bibliography{eGaIn}

\clearpage
\section{Supplementary Information}

\setcounter{figure}{0}
\renewcommand{\figurename}{Figure}
\renewcommand{\thefigure}{S\arabic{figure}}

\subsection{Sweep rate dependence}

\begin{figure}[h]
	\centering
	\includegraphics[width=0.6\columnwidth]{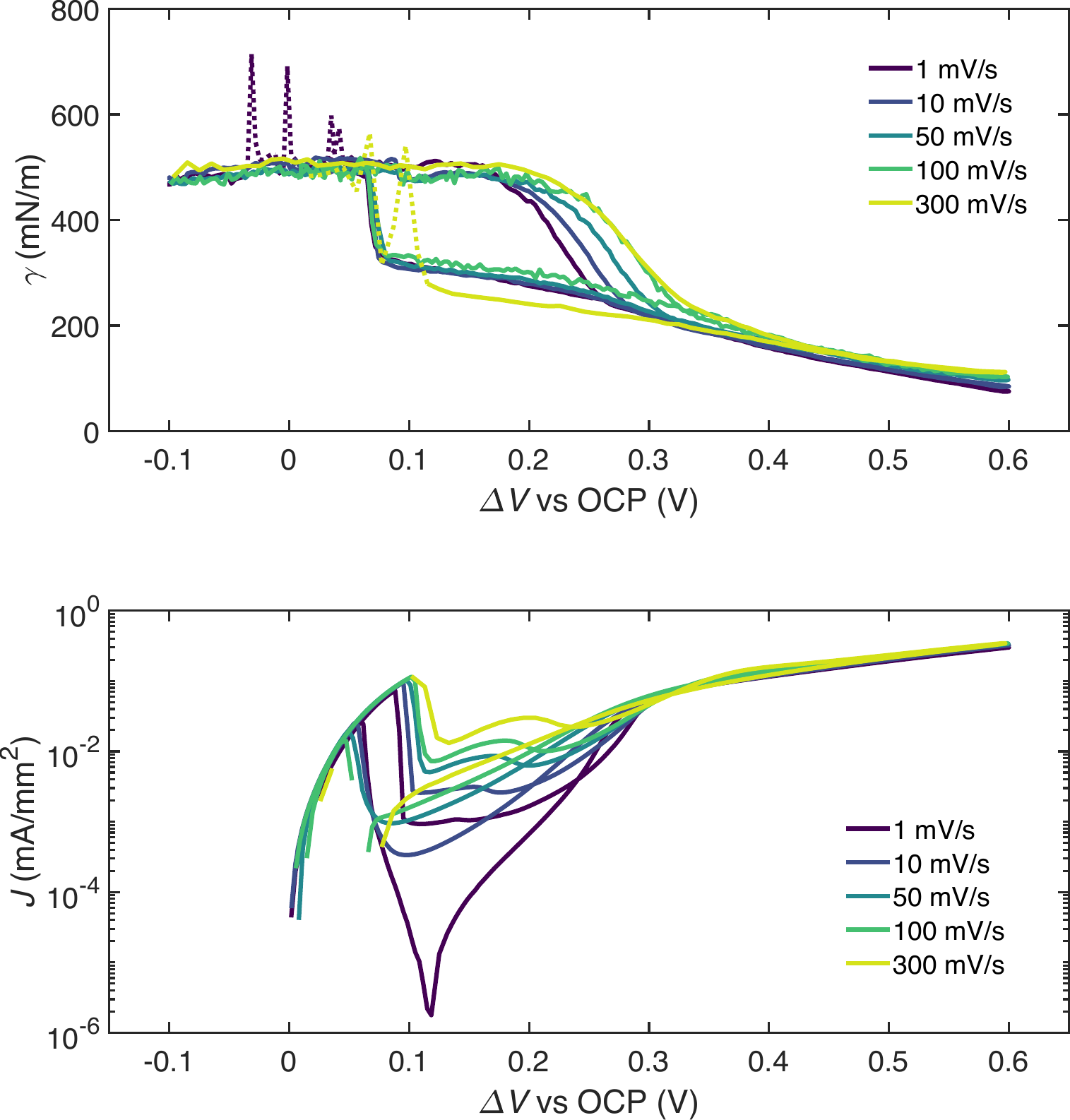
	}
\caption{Effect of cyclic voltammetry scan rate on the degree of hysteresis, measured via both the interfacial tension $\gamma$ and the current density $J$.  Dashed lines represent measurements that were affected by either air bubbles sticking to the droplet or for the presence of significant oscillations. These measurements were done using a copper counter electrode (rather than platinum).}
	\label{f:sweeprate}
\end{figure}

Figure~\ref{f:sweeprate} shows how the degree of hysteresis depends on the voltage sweep rate, via cyclic voltammetry experiments performed at 1 mV/s, 10 mV/s, 50 mV/s, 100 mV/s, and 300 mV/s. In the forward sweeps, we observe that both $V_\mathrm{SQ}$ and the transition from the transpassive to steady state regime move to higher voltages for higher scan rates, while the reverse sweeps are less rate-dependent, as observed most clearly in the interfacial tension $\gamma$ measurements. Once the system is in the steady state regime, both $\gamma$ and $J$ are largely rate-independent.  From the  current density $J$ measurements, we also observe that both the initial oxidation peak and the second smaller oxidation peak slightly increase in both $J$ and $\Delta V$.  In the reverse sweeps, the fastest two sweep rates exhibit reduction peaks, while the slower scan rates have oxidation peaks.  Note that $\gamma$ varies slightly in the reverse sweep of the 100 mV/s, and greatly in the 300 mV/s. This may arise from the inertia of the droplet resisting changes in shape at these fastest rates: the fluid motion cannot keep pace with the rapid changes in interfacial tension. In order to seek a regime in which hysteresis is no longer present, it would be necessary to explore sweep rates far lower than 1 mV/s; the long time scales involved make this regime experimentally-challenging.

\subsection{Image analysis pipeline}

\begin{figure}[h]
    \centering
    \includegraphics[width=0.8\columnwidth]{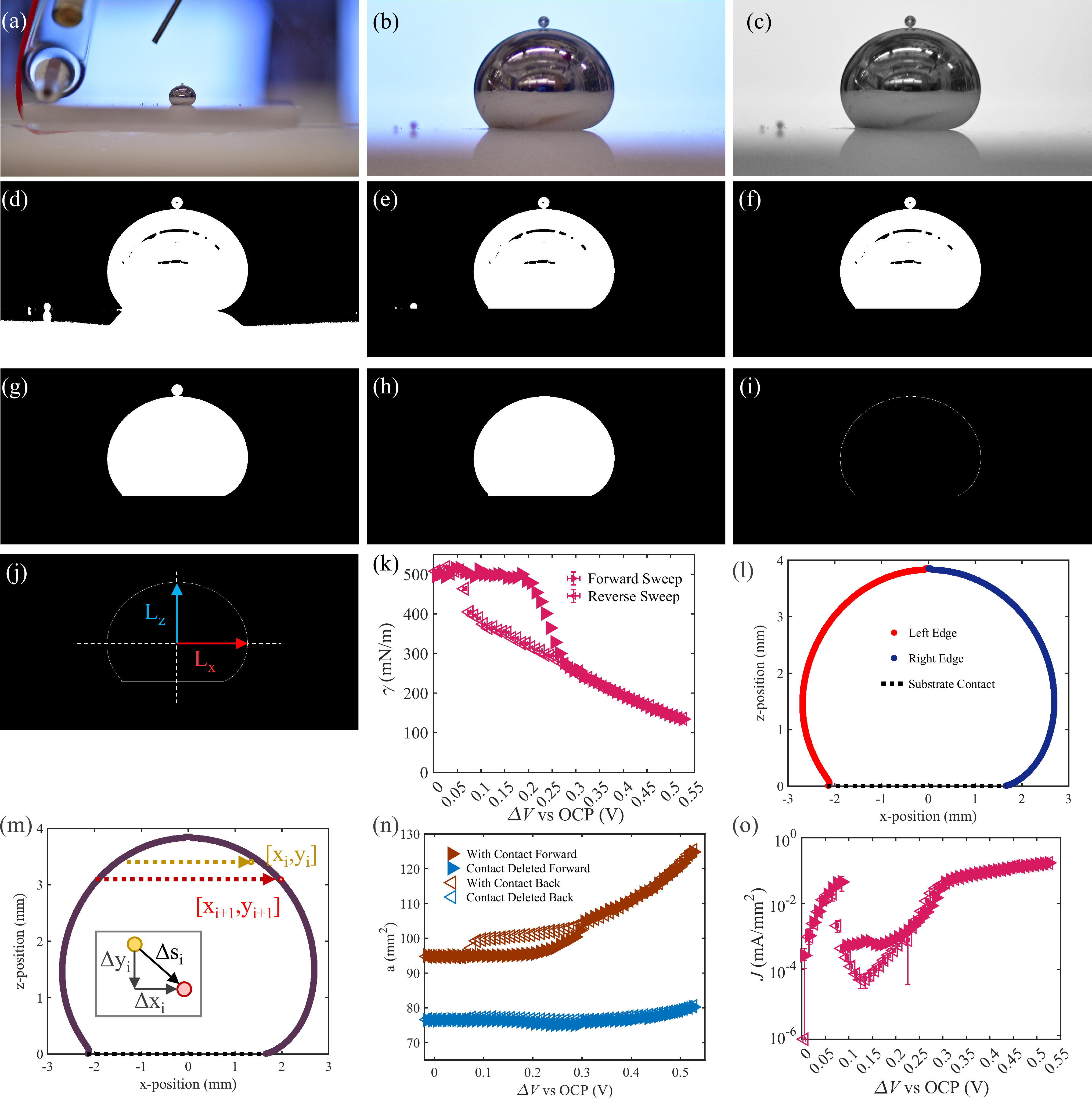}
    \caption{Sessile droplet image analysis pipeline. Original camera image (a), cropped (b), converted to grayscale (c), and binarized (d). The image is cleaned by removing pixels below the contact line (e), isolating the central droplet (f), closing interior holes (g), and using a Hough transform to find the largest droplet for analysis. We use edge-detection isolate the boundary (i) of the droplet, and measure  the equatorial radius $L_x$, and distance from the top of the droplet to the equator $L_z$ (j).  These two values measure the interfacial tension via Equation~\ref{eq:hutzler}, plotted for all values of $\Delta V$ in (k). To extract the surface area, we split the droplet at its centerline (l), recenter it around $x=0$, and numerically integrate (m) to find the approximate surface area $a$, assuming it is radially symmetric. The area $a$ of the droplet changes as it flattens (n),  but the exposed surface area $a$ remains approximately constant ($\approx (77  \pm 1)\, \mathrm{mm}^2$). This value of $a$ is used to calculate the current density, $J = I/a$ (o).}
    \label{f:DropArea}
\end{figure}

\subsection{Dynamics without an applied potential}

\begin{figure}[h]
  \centering
  \includegraphics[width=0.4\columnwidth]{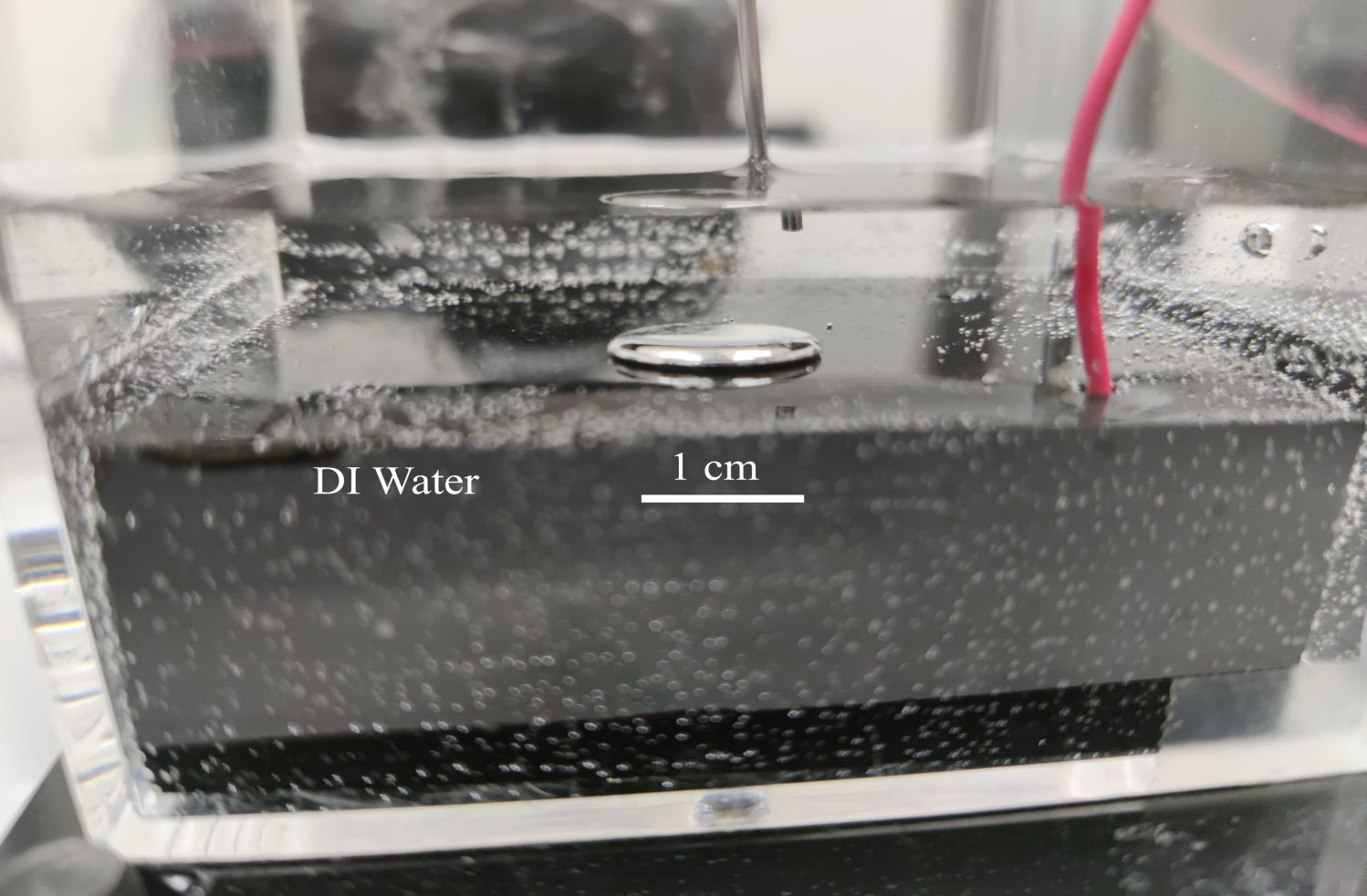}
  \caption{A 100 $\mu$L droplet of EGaIn flattened after  sitting in dionized water overnight. No electric potential has been applied to the droplet.}
  \label{f:DIWater}
\end{figure}

We include some  observations about two related systems for which the oxides {\it are} known. EGaIn in placed in dionized water is known to form a thick GaOOH shell, and correspondingly decrease $\gamma$ \cite{khan_influence_2014}. This is shown in Figure~\ref{f:DIWater}. In contrast, a droplet of EGaIn in air forms a shell of Ga$_\mathrm{2}$O$_\mathrm{3}$, though it is not observed to flatten into a disk due to the shell being strong enough to hold the shape of the droplet under gravitational pressure. 

\subsection{Kramer-Kronig test on electrochemical impedance data}

We collected both low  to high frequency (L2H) sweep data and the more standard  high to low frequency sweep (H2L), in order to better establish the reliability of the data in the face of known hysteretic effects. 
The Nyquist data for both H2L and L2H were modeled using the circuits shown in Figure~\ref{f:EISCircuits}, and the fits are shown in Figure~\ref{f:EISFitData}, with the capacitance and the resistance of the interface determined by the fits shown in Figure~\ref{f:EISFits}. 
While the trends are similar for both the H2L and L2H data, we do note that  at 0.1 V the magnitude of the impedance values measured for the H2L data is larger than the L2H data, which is likely due to high oxidation at the beginning of the L2H sweep compared to the H2L sweep.
Additionally, the variance in the data points in the L2H sweep at the 0.25 V bias is likely due to the inductance behavior of the liquid metal interface as it undergoes the formation of multiple unstable oxide species during the low frequency measurement. One of the oxide species became stable at the higher frequencies, reducing this effect.

Figure~\ref{f:EISall} shows the Kramer-Kronig (KK) model fit for both direction sweeps, in which we observed that the EIS data satisfies the KK test within error. We do see slight deviations at high frequencies for both H2L and L2H data collected at  DC bias values  0.05 V, 0.40 V and 0.50 V; this is  not unexpected in this type of system  due to an accumulation of oxide species near the electrode interface from the low frequency time scale. These species could form an additional charged interface, especially at the potential where the surface oxide species is stable.
We considered the possibility that effects (all at high frequencies) could be due to noise artifacts, but the similar resistive trend observed in both the H2L and L2H Nyquist plots suggests the of presence of a charged interface or a thin oxide film that was not accounted for in our model.

\begin{figure}
  \centering
  \includegraphics[width=\columnwidth]{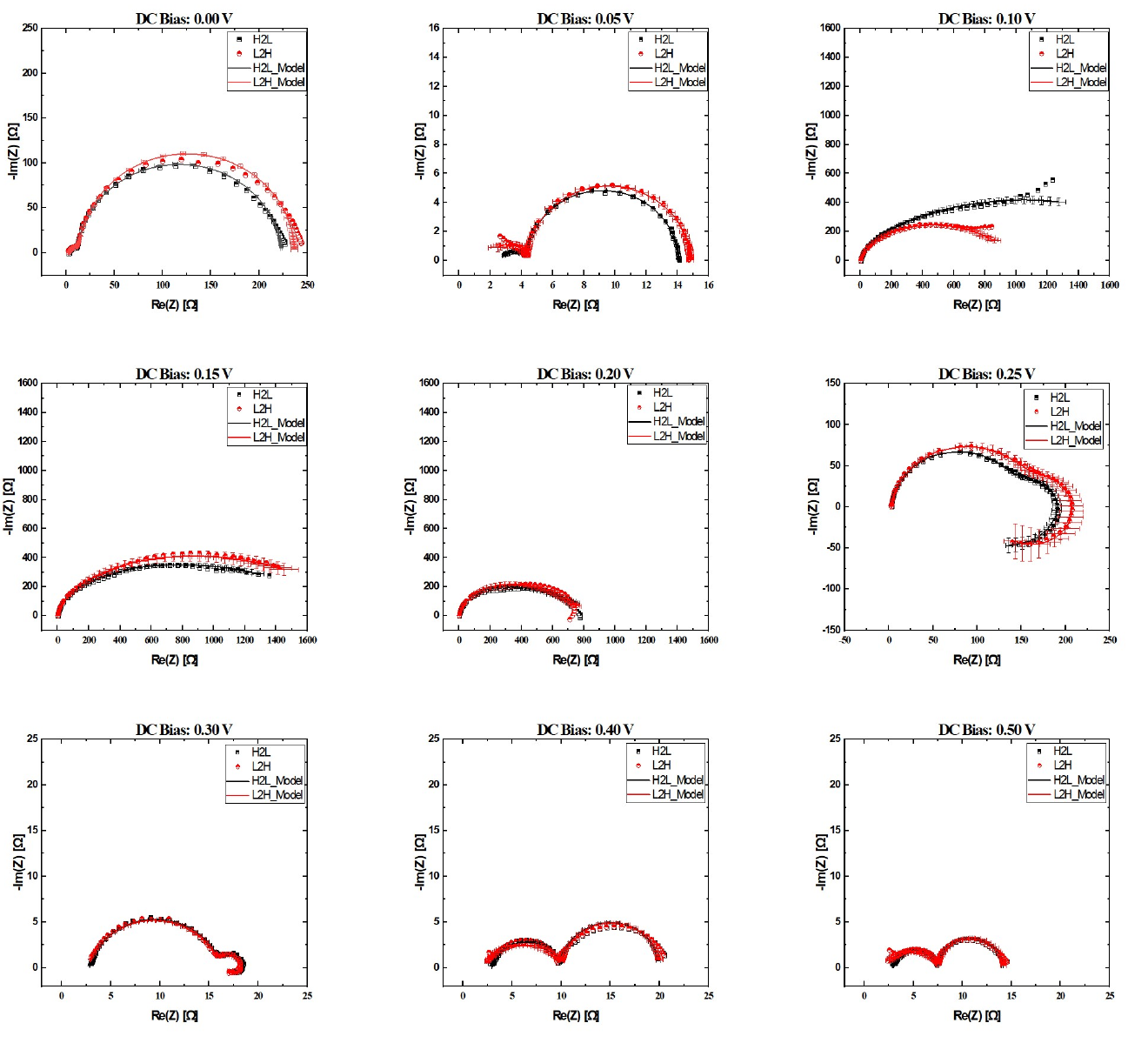}
  \caption{Nyquist plots fit to their respective circuit models at different oxidative potentials for all regimes  (different DC bias) in forward sweep. The filled triangles show the measured data, while the unfilled triangles show the circuit model fits, with the black upwards facing triangles showing the high to low frequency (H2L) sweep, and the downward red facing triangles showing the low to high frequency (L2H) sweep.}
  \label{f:EISFitData}
\end{figure}

\begin{figure}
  \centering
  \includegraphics[width=\columnwidth]{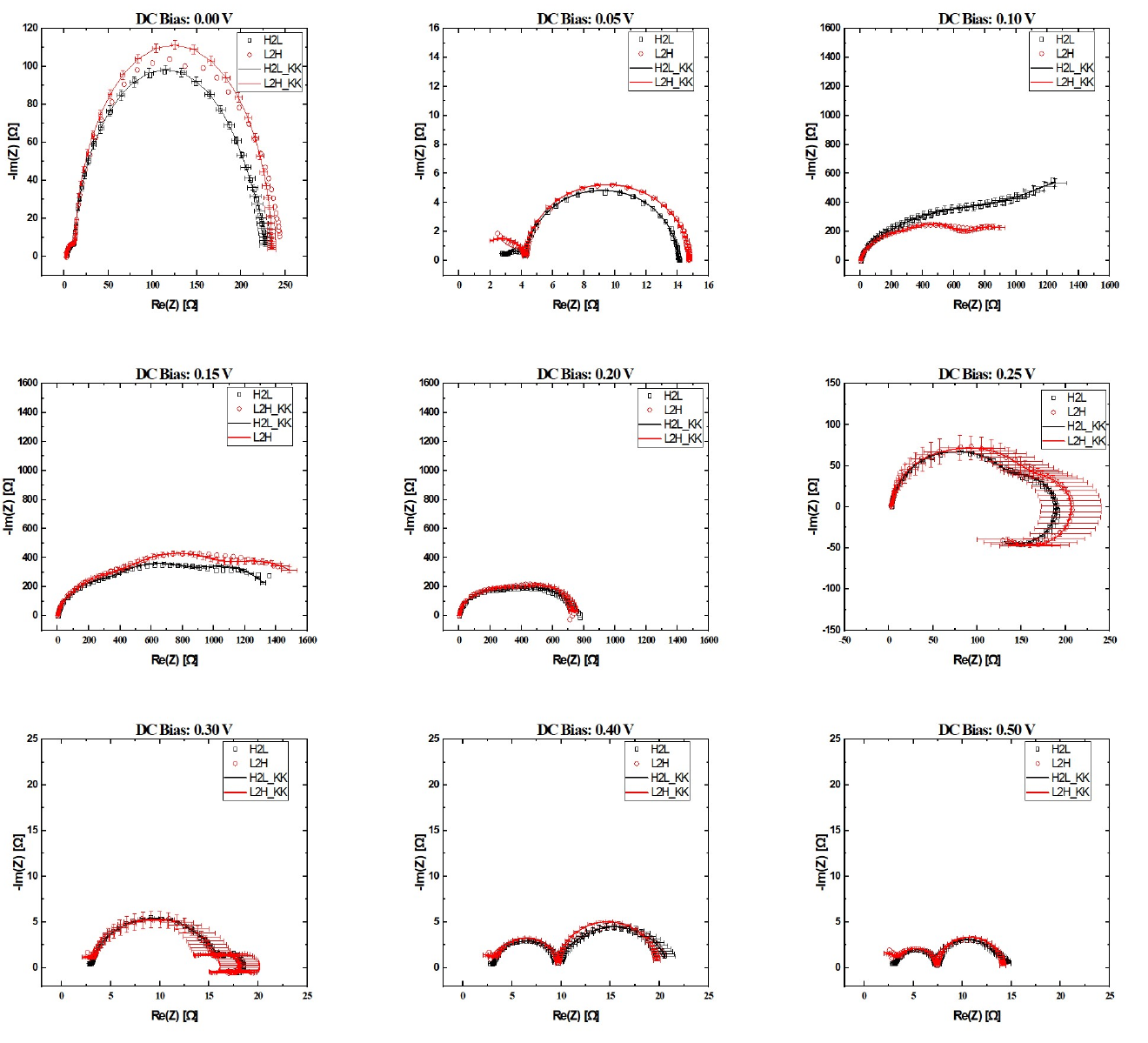}
  \caption{Nyquist plots and their respective KK tests at different oxidative potentials for all regimes  (different DC bias) in forward sweep. The filled triangles show the measured data, while the unfilled triangles show the KK fits, with the black upwards facing triangles showing the high to low frequency (H2L) sweep, and the downward red facing triangles showing the low to high frequency (L2H) sweep.}
  \label{f:EISall}
\end{figure}

\end{document}